\documentstyle[aps,prl,twocolumn]{revtex}

\draft

\def\lesssim{\mathrel{\mathpalette\vereq<}}

\makeatletter
\def\vereq#1#2{\lower3pt\vbox{\baselineskip1.5pt \lineskip1.5pt
\ialign{$\m@th#1\hfill##\hfil$\crcr#2\crcr\sim\crcr}}}
\makeatother

\begin{document}

\preprint{LBNL-42967, UCB-PTH-99/07}

\title{Can $\epsilon'/\epsilon$ be supersymmetric?\cite{thankyou}}

\author{Antonio Masiero}
\address{SISSA, Via Beirut 2-4, 34014 Trieste, Italy}
\address{INFN, Sezione di Trieste, Italy}
\author{Hitoshi Murayama}
\address{Department of Physics, University of California, Berkeley, CA
  94720}
\address{Theoretical Physics Group, Lawrence Berkeley National 
Laboratory, Berkeley, CA 94720}

\maketitle

\begin{abstract}
The possible supersymmetric contribution to $\epsilon'/\epsilon$ has 
been generally regarded small in the literature.  We point out, however, that 
this is a result based on specific assumptions, such as universal 
scalar mass, and in general needs not to be true.  Based on a general situation,
(1) hierarchical quark Yukawa matrices protected by flavor symmetry, 
(2) generic dependence of Yukawa matrices on Polonyi/moduli fields as 
expected in many supergravity/superstring theories, (3) Cabibbo 
rotation originating from the down-sector, and (4) phases of order 
unity, we find the typical supersymmetric contribution to 
$\epsilon'/\epsilon$ to be order $3 \times 10^{-3}$ for $m_{\tilde{q}} 
= 500$~GeV. It is even possible that the supersymmetric contribution 
dominates in the reported KTeV value $\epsilon'/\epsilon = (28 \pm 
4.1)\times 10^{-4}$.  If so, the neutron electric dipole moment is 
likely to be within the reach of the currently planned experiments.
\end{abstract}

\pacs{~}

\narrowtext

CP violation is the least understood aspect in the properties of the
fundamental particles besides the mechanism of the electroweak
symmetry breaking.  The so-called ``indirect CP violation'' $\epsilon$
in the
neutral kaon system has been known for three decades as the only
evidence that there is a fundamental distinction between particles and
anti-particles.  This year, however, produced two new manifestations of
CP violation: $\sin 2\beta$ from $B \rightarrow \psi K_{s}$ at CDF
\cite{CDF}, even though the evidence is still somewhat week, and a
beautiful measurement of ``direct CP violation'' $\epsilon'/\epsilon$
in neutral kaon system from KTeV \cite{KTeV}. The latter confirmed the
previous evidence reported by NA31 \cite{NA31} at a much higher accuracy and
excludes the so-called superweak model of CP violation.  The reported
number, $\epsilon'/\epsilon = (28 \pm 4.1)\times 10^{-4}$ \cite{KTeV}
was, however, somewhat surprisingly large.  The standard model
prediction is currently controversial (see
Table~\ref{tab:estimates}) and is dominated by theoretical
uncertainties in quantities such as the non-perturbative matrix
elements and
the strange quark mass $m_{s}$.  Given this situation, one cannot
interpret the KTeV data reliably; in particular, it is not clear if
the data is consistent with the standard model (see also for a recent
discussion in \cite{Sanda}).

On the other hand, the standard model is believed to be only an 
effective low-energy approximation of fundamental physics.  This is 
largely because it lacks a dynamical explanation of the mechanism of 
electroweak symmetry breaking and suffers from a serious hierarchy 
problem that the electroweak scale is unstable against radiative 
corrections.  The best available simultaneous solution to both of 
these problems is supersymmetry.  Therefore, it is a natural question 
to ask if supersymmetry gives a sizable contribution to 
$\epsilon'/\epsilon$ given a precise measurement.  The experimental 
sensitivity to a possible supersymmetric contribution is currently 
plagued by the theoretical uncertainties mentioned above, but we can 
expect them to be resolved or at least alleviated eventually by
improvements in particular in  lattice QCD 
calculations.  It is therefore timely to reconsider the supersymmetric 
contribution to $\epsilon'/\epsilon$.

In this letter, we revisit the estimate of $\epsilon'/\epsilon$ in 
supersymmetric models \cite{Nir}.  The common lore in the literature is that 
the supersymmetric contribution to $\epsilon'/\epsilon$ is in general 
rather small.  We point out, however, that this lore is largely based 
on the specific choice of supersymmetry breaking effects sometimes 
called minimal supergravity framework \cite{GG}.  A more general framework of 
flavor structure tends to give a relatively large contribution to 
$\epsilon'/\epsilon$ in a wide class of models.  The assumptions are: 
(1) hierarchical quark Yukawa matrices protected by flavor symmetry, 
(2) generic dependence of Yukawa matrices on Polonyi/moduli fields as 
expected in many supergravity/superstring theories, (3) Cabibbo 
rotation originating from the down-sector, and (4) phases of order 
unity.  In fact, there is even an intriguing possibility that the 
observed $\epsilon'/\epsilon$ is mostly or entirely due to the 
supersymmetric contribution.

To discuss the CP violating effects induced by loops of supersymmetric 
particles, it is convenient to introduce the mass insertion formalism
\cite{Hall}.  The Yukawa matrices are couplings in the superpotential $W = 
Y^{u}_{ij} Q_{i} U_{j} H_{u} + Y^{d}_{ij} Q_{i} D_{j} H_{d}$, where 
$H_{d}$, $H_{u}$ are Higgs doublets and $i,j$ flavor indices.  
The expectation values $\langle H_{u} \rangle = v 
\sin \beta/\sqrt{2}$ and $\langle 
H_{d} \rangle = v \cos\beta/\sqrt{2}$ 
generate quark mass matrices $M^{u} = Y^{u} v 
\sin\beta/\sqrt{2}$ and $M^{d} = Y^{d} v \cos\beta/\sqrt{2}$.  They 
are diagonalized by bi-unitary transformations $M^{u} = V^{u*}_{L} 
{\rm diag}(m_{u}, m_{c}, m_{t}) V^{u\dagger}_{R}$ and $M^{d} = 
V^{d*}_{L} {\rm diag}(m_{d}, m_{s}, m_{b}) V^{d\dagger}_{R}$, and the 
Cabibbo--Kobayashi--Maskawa matrix is given by $V^{u\dagger}_{L} 
V^{d}_{L}$.  The squarks have chirality-preserving mass-squared 
matrices ${\cal L} \supset - \tilde{Q}_{i}^{*} (M^{2}_{Q})_{ij} 
\tilde{Q}_{j} - \tilde{U}_{i}^{*} (M^{2}_{U})_{ij} \tilde{U}_{j} - 
\tilde{D}_{i}^{*} (M^{2}_{D})_{ij} \tilde{D}_{j}$ and 
chirality-violating trilinear couplings ${\cal L} \supset - 
\tilde{Q}_{i} (A^{d})_{ij} \tilde{D}_{j} H_{d} - \tilde{Q}_{i} 
(A^{u})_{ij} \tilde{U}_{j} H_{u}$ where $H_{d}$, $H_{u}$ are Higgs 
doublets.  The Higgs expectation values 
generate the left-right (LR) mass-squared matrix $M^{2,d}_{LR} = 
A^{d} v \sin\beta/\sqrt{2}$ and $M^{2,u}_{LR} = A^{u} v 
\cos\beta/\sqrt{2}$.  The convenient basis to discuss flavor-changing 
effects in the gluino loop diagrams is the so-called superCKM basis
\cite{Hall}. In this basis  the relevant quark mass 
matrix is diagonalized (say, $M^{d}$) and the squarks are also rotated 
in the same way, $M^{2}_{Q} \rightarrow V^{d\dagger}_{L} M^{2}_{Q} 
V^{d}_{L}$, $M^{2}_{D} \rightarrow V^{d\dagger}_{R} M^{2}_{D} 
V^{d}_{R}$, and $M^{2,d}_{LR} \rightarrow {}^t V^{d}_{L} 
M^{2,d}_{LR} V^{d}_{R}$.  Flavor-changing effects can be estimated by 
insertion of flavor-off-diagonal components of the mass-squared 
matrices in this basis.  By normalizing the off-diagonal components by 
average squark mass-squared $m^{2}_{\tilde{q}}$, we define 
$(\delta^{d}_{LL})_{ij} = (V^{d\dagger}_{L} M^{2}_{Q} 
V^{d}_{L})_{ij}/m^{2}_{\tilde{q}}$, $(\delta^{d}_{RR})_{ij} = 
(V^{d\dagger}_{R} M^{2}_{D} V^{d}_{R})_{ij}/m^{2}_{\tilde{q}}$, and 
$(\delta^{d}_{LR})_{ij} = ({}^t V^{d}_{L} M^{2,d}_{LR} 
V^{d}_{R})_{ij}/m^{2}_{\tilde{q}}$.

The supersymmetric contributions due to gluino loops to neutral Kaon 
parameters $(\Delta m_{K})_{SUSY}$, $\epsilon_{SUSY}$ and 
$(\epsilon'/\epsilon)_{SUSY}$ have been calculated, and have been used 
to place bounds on mass insertion parameters \cite{Hagelin}.  The 
values of the mass insertion parameters 
which saturate the observed numbers of $\Delta m_{K}$, $\epsilon$ 
and $\epsilon'/\epsilon$ are tabulated in Table~\ref{tab:bounds}, 
after updating the numbers in Ref.~\cite{Gabbiani}.  These 
numbers are subject to theoretical uncertainties in QCD corrections and 
matrix elements at least of order a few tens of percents (this is at
least what is obtained for the $\Delta S=2$ transitions 
\cite{Bagger}).  Barring  possible cancellations with the 
standard-model amplitudes as well as with the other SUSY contributions
({\it i.e.}\/, chargino and charged Higgs exchages), the mass
insertion parameters have to
be  smaller than or at most comparable to the entries in the Table.  
Stringent bounds on $(\delta^{d}_{12})_{LL}$ from $\Delta m_{K}$ and $\epsilon$ 
have been regarded as a problem in supersymmetric models.  A random 
mass-squared matrix of squarks would lead to a large 
$(\delta^{d}_{12})_{LL}$ which overproduce $\Delta m_{K}$ or 
$\epsilon$.  Usually 
an assumption is invoked that the squark mass-squared matrix is 
proportional to the identity matrix (universality), at least for the 
first- and second-generations (alternatively one can invoke an
alignment of the quark and squark mass matrices \cite{alignment}). Even when 
such an assumption is made at the Planck scale, radiative effects can 
induce $(\delta^{d}_{12})_{LL}$ and hence over produce $\Delta m_{K}$ or 
$\epsilon$.  Once the bounds are satisfied, however, the 
supersymmetric contribution to $\epsilon'$ tends to be rather small: 
$\Delta m_{K}$ and $\epsilon$ require $|(\delta^{d}_{12})_{LL}| = 
(({\rm Re}(\delta^{d}_{12})_{LL}^{2})^{2} +({\rm 
Im}(\delta^{d}_{12})_{LL}^{2})^{2})^{1/4} \lesssim 0.019$--0.092, 
which is much smaller than the corresponding bounds from 
$\epsilon'/\epsilon$, $|{\rm Im}(\delta^{d}_{12})_{LL}| \lesssim 
0.10$--0.27 \cite{foot}. This fact led to a 
common wisdom that the supersymmetric contribution to $\epsilon'$ is 
in general small.  For the rest of the letter, we simply assume 
that $(\delta^{d}_{ij})_{LL}$ parameters are under control by some 
mechanisms such as a flavor symmetry and do not go into any further 
discussions.

However, the contribution from $(\delta^{d}_{12})_{LR}$ can be 
important; even $|{\rm Im}(\delta^{d}_{12})_{LR}^{2}| \sim 10^{-5}$ 
gives a significant contribution to $\epsilon'/\epsilon$ while the 
bounds on $(\delta^{d}_{12})_{LR}$ from $\Delta m_{K}$ and $\epsilon$ 
are only about $3 \times 10^{-3}$ and $3 \times 10^{-4}$, 
respectively.  Actually, one can even imagine to saturate both 
$\epsilon$ and $\epsilon'/\epsilon$ at the borderline of the current 
limits \cite{Silvestrini}.  Therefore, whether the supersymmetric 
contribution to $\epsilon'/\epsilon$ can be important is an issue of 
how large $(\delta^{d}_{12})_{LR}$ is expected in supersymmetric models 
rather than that of phenomenological viability.

What is the general expectation on the size of 
$(\delta^{d}_{12})_{LR}$?  The common answer in the literature to this 
question is that it is very small in general, and hence the 
supersymmetric contribution to $\epsilon'$ has been regarded small as 
well.  This is indeed the case if one assumes that all soft 
supersymmetry breaking parameters are universal at Planck- or 
GUT-scale where $(\delta^{d}_{12})_{LR}$ is induced only radiatively 
at higher orders in small Yukawa coupling constants of first and 
second generation particles \cite{GG}.  However, the universal breaking is a 
strong assumption and is known not to be true in many supergravity and 
string-inspired models \cite{BIM}.  On the other hand, the LR mass matrix has
the  same flavor structure as the fermion Yukawa matrix and both in fact 
originate from the superpotential couplings.  Our theoretical 
prejudice is that there is an underlying symmetry (flavor symmetry) 
which restricts the form of the Yukawa matrices to explain their 
hierarchical forms.  Then the LR mass matrix is expected to have a 
very similar form as the Yukawa matrix.  More precisely, we expect the 
components of the LR mass matrix to be roughly the supersymmetry 
breaking scale ({\it e.g.}\/, $m_{3/2}$) times the corresponding 
component of the quark mass matrix.  However, there is no reason for them 
to be simultaneously diagonalizable based on this general argument.  
In general, we expect the size of $(\delta^{d}_{12})_{LR}$ to be
\begin{equation}
  (\delta^{d}_{12})_{LR} \sim \frac{m_{3/2} M^{d}_{12}}{m_{\tilde{q}^{2}}}.
\end{equation}

To be more concrete, one can imagine a string-inspired theory where 
the Yukawa couplings in the superpotential $W = Y^{d}_{ij} (T) Q^{i} 
D^{j} H_{d}$ are in general complicated functions of the moduli fields 
$T$.  The moduli fields have expectation values of order string scale 
$\langle T \rangle$ which describe the geometry of compactified extra 
six dimensions.  The low-energy Yukawa couplings are then given by 
their expectation values $Y^{d}_{ij}(\langle T \rangle)$.  On the 
other hand, the moduli fields in general also have couplings to fields 
in the hidden sector and acquire supersymmetry-breaking $F$-component 
expectation values $F_{T} \sim m_{3/2}$ in the Planck unit $M_{Pl} = 
\sqrt{8\pi}$.  This generates trilinear couplings given by \cite{BIM}
\begin{equation}
        {\cal L} \supset \frac{\partial Y^{d}_{ij}}{\partial T}\langle F_{T} 
        \rangle \tilde{Q}^{i} \tilde{D}^{j} H_{d},
\end{equation}
which depend on a different matrix $\partial Y^{d}_{ij}/\partial T$.  
Due to holomorphy, flavor symmetry is likely to constrain $Y^{d}_{ij}$ 
and its derivative to be similar while they in general do not have to 
exactly proportional to each other and hence are not simultaneously 
diagonalizable.

In order to proceed to numerical estimates of $(\delta^{d}_{12})_{LR}$, we 
need to specify if the quark mixings come from up- or down-sector.  In 
general, attributing mixing to up-sector gives smaller flavor-changing 
effects and receive less constraints \cite{alignment}.  
On the other hand, historically 
the Cabibbo angle has been often attributed to the down sector because 
of a numerical coincidence $V_{us} = \sin \theta_{C} = 0.22 \sim 
\sqrt{m_{d}/m_{s}}$.  For our purpose, we pick the latter choice, 
which fixes the form of the mass matrix for the first and second 
generations to be
\begin{equation}
        M^{d} \simeq \left( \begin{array}{cc}
                m_{d} & m_{s} V_{us} \\
                 & m_{s}
        \end{array} \right),
\end{equation}
where the (2,1) element is unknown due to our lack of knowledge on 
the mixings among right-handed quarks.  Based on the general 
considerations on the LR mass matrix above, we expect
\begin{equation}
        m^{2,d}_{LR} \simeq m_{3/2}  \left( \begin{array}{cc}
                a m_{d} & b m_{s} V_{us} \\
                 & c m_{s}
        \end{array} \right) ,
\end{equation}
where $a$, $b$, $c$ are constants of order unity.  Unless $a=b=c$ 
exactly, $M_{d}$ and $m^{2,d}_{LR}$ are not simultaneously 
diagonalizable and we find
\begin{eqnarray}
  \lefteqn{
        (\delta^{d}_{12})_{LR} \simeq \frac{m_{3/2} m_{s} 
        V_{us}}{m_{\tilde{q}}^{2}} } \nonumber \\
    & & = 2 \times 10^{-5}
        \left( \frac{m_{s}(M_{Pl})}{\rm 50~MeV}\right)
        \left(\frac{m_{3/2}}{m_{\tilde{q}}}\right)
        \left(\frac{\rm 500~GeV}{m_{\tilde{q}}}\right).
        \label{eq:estimate}
\end{eqnarray}
It is interesting to see that $(\delta^{d}_{12})_{LR}$ of this naive 
dimensional estimate gives the saturation of the bound from 
$\epsilon'/\epsilon$ (see Table~\ref{tab:bounds}) if it has a phase of 
order unity.  

The key-point of the above example is that the large
value of $\epsilon'/\epsilon$ of the KTeV and NA31 experiments can be
accounted for in the supersymmetric context without particularly
contrived assumptions on the size of the $(\delta^{d}_{12})_{LR}$ mass
insertion  with the exception of taking it to have a large CP violating
phase \cite{Isidori}.

One may wonder if typical off-diagonal
elements in  $(\delta^{d}_{ij})_{LR}$ may be already excluded from other 
flavor-changing processes.  For instance, $(\delta^{d}_{23})_{LR}$ is 
constrained by $b\rightarrow s\gamma$ to be less than 1--3$\times 10^{-2}$
\cite{Gabbiani}.  This is to be compared to the estimate 
$(\delta^{d}_{23})_{LR} \sim m_{3/2} m_{b} 
V_{cb}/m_{\tilde{q}}^{2} \sim 2 \times 10^{-5}$.  The constraint 
from $b\rightarrow s l^{+} l^{-}$ is similarly insignificant 
\cite{Lunghi}.  Constraints from the up sector are much weaker.

It is tempting to speculate that the observed $\epsilon'/\epsilon$ may
be dominated by $(\delta^{d}_{12})_{LR}$ contribution.  The estimate
in Eq.~(\ref{eq:estimate}) requires $m_{3/2} \sim m_{\tilde{q}}$ and
an $O(1)$ phase.  Because $m_{\tilde{q}}^{2}$ acquires a positive
contribution from gluino mass in the renormalization-group evolution
while off-diagonal components in LR mass matrix don't, such a scenario
would prefer models where the gaugino mass is somewhat smaller than
scalar masses (assumed also to be $O(m_{3/2})$).  An important implication
of supersymmetry-dominated $\epsilon'/\epsilon$ is that neutron
electric dipole moment (EDM) is likely to be large.  The current limit
on neutron EDM $d_{n} < 11 \times 10^{-26} e{\rm cm}$ constrains
$|{\rm Im}(\delta^{d}_{11})_{LR}| < (2.4, 3.0, 5.6) \times 10^{-6}$ for
$m^2_{\tilde{g}}/m^2_{\tilde{q}} = 0.3, 1.0, 4.0$, respectively, with
a theoretical uncertainty of at least a factor of two, while our
estimate gives $(\delta^{d}_{11})_{LR} \sim m_{3/2}
m_{d}/m_{\tilde{q}}^{2} \sim 3 \times 10^{-6}$.  It would be
interesting to see results from near-future experiments which are
expected to improve the limit on $d_{n}$ by two orders of magnitude.

One may extend the discussion to the lepton sector.  Let us  consider
$m_{\tilde{l}} \sim m_{3/2} \sim 500$~GeV for our discussions.  
The constraints from $\mu \rightarrow e \gamma$ and the electron EDM are:
$|(\delta^{l}_{12})_{LR}| < 0.7$--$1.9 \times 10^{-5}$ 
and $|{\rm Im}(\delta^{l}_{11})_{LR}| < 1.5$--$3.5 \times 10^{-6}$ for 
$0.4 < m_{\tilde{\gamma}}^{2}/m_{\tilde{l}}^{2} < 5.0$ \cite{Gabbiani}.  
Our estimates on these mass insertion parameters are 
$(\delta^{l}_{12})_{LR} \sim m_{3/2} m_{\mu} V_{\nu_{e}
  \mu}/m_{\tilde{l}}^{2} \sim 2.1 \times 10^{-4} V_{\nu_{e} \mu}$ 
  and $(\delta^{l}_{11})_{LR} \sim m_{3/2} m_{e}/m_{\tilde{l}}^{2} \sim
1.0 \times 10^{-6}$.  In the lack of our knowledge on the lepton
mixing angles, we cannot draw a definite conclusion on the $\mu \rightarrow
e\gamma$ process.  One possible choice is what is suggested by the 
small angle MSW solution to the solar neutrino problem, $V_{\nu_{e} 
\mu} \sim \sqrt{m_{e}/m_{\mu}} \sim 0.05$.
It is interesting that these estimates of $|(\delta^{l}_{12})_{LR}|$ 
and $|{\rm Im}(\delta^{l}_{11})_{LR}|$ nearly saturate the bounds.

In summary, we have reconsidered the possible supersymmetric contribution 
to $\epsilon'/\epsilon$.  Contrary to the lore in the literature, we 
find that generic supersymmetric models give an interesting contribution
to $\epsilon'/\epsilon$, and it is even possible that it dominates
in the observed value.  We expect the neutron EDM to be within the 
reach of near-future experiments in that case.

\acknowledgements
A.M. thanks Luca Silvestrini for interesting discussions and
suggestions. 
H.M. thanks Kaustubh Agashe and Lawrence Hall for pointing out
numerical errors in an earlier version of the paper.
We thank the organizers of the IFT workshop, ``Higgs and
SuperSymmetry:
Search \& Discovery,''  University of Florida, March 8-11, 1999 for
providing the stimulating atmosphere for this work to be started.

\begin{table}
\caption{Various estimates of $\epsilon'/\epsilon$ in the literature.
  Two Buras' estimates use different values of the strange quark mass, 
 $m_s (2~{\rm GeV}) = 130 \pm 20~{\rm MeV}$ (QCD sum rule) and $110
 \pm 20~{\rm MeV}$ (lattice QCD).}
\label{tab:estimates}
\begin{center}
\begin{tabular}{lc}
Reference & $\epsilon'/\epsilon$ \\ \hline
Ciuchini \cite{Ciuchini} & $(4.6 \pm 3.0 \pm 0.4) \cdot 10^{-4}$\\
Buras \cite{Buras} & $(5.7 \pm 3.6) \cdot 10^{-4}$ \\
Buras \cite{Buras} & $(9.1 \pm 5.7) \cdot 10^{-4}$ \\
Bertolini \cite{Bertolini} & $(17 ^{+14}_{-10}) \cdot 10^{-4}$ \\
\end{tabular}
\end{center}
\end{table}

\begin{table}
  \caption{The values of the mass insertion parameters which saturate 
    the observed numbers
    $(\Delta m_{K})_{SUSY} = 3.521 \times 10^{-12}$~MeV,~$(\epsilon)_{SUSY} = 2.27 \times 10^{-3}$, and  
    $(\epsilon'/\epsilon)_{SUSY} = 2.8 \times 10^{-3}$ for three 
    values of $x \equiv m_{\tilde{g}}^{2}/m_{\tilde{q}}^{2}$.  They 
    are based on Ref.~\protect\cite{Gabbiani} scaled to the current
    observed numbers.  The
    squark mass is taken to be 500~GeV, and have theoretical
    uncertainties of at least a few tens of percents.  Barring possible 
    cancellations, the mass insertion parameters must be smaller than 
    or at most comparable to the entries given here.}
  \label{tab:bounds}
  \begin{tabular}{ccccc}
    \noalign{\smallskip}
    $\Delta m_{K}$ & $x$ 
    & $|{\rm Re}(\delta^{d}_{12})_{LL}^{2}|^{\frac{1}{2}}$
    & $|{\rm Re}(\delta^{d}_{12})_{LR}^{2}|^{\frac{1}{2}}$
    & $|{\rm Re}(\delta^{d}_{12})_{LL}(\delta^{d}_{12})_{RR}|^{\frac{1}{2}}$\\
    \noalign{\smallskip} \hline \noalign{\smallskip}
    & 0.3 & 0.019 & 0.0078 & 0.0025\\
    & 1.0 & 0.040 & 0.0044 & 0.0028\\
    & 4.0 & 0.092 & 0.0053 & 0.0040\\
    \noalign{\smallskip} \hline \noalign{\smallskip}
    $\epsilon$ & $x$ 
    & $|{\rm Im}(\delta^{d}_{12})_{LL}^{2}|^{\frac{1}{2}}$
    & $|{\rm Im}(\delta^{d}_{12})_{LR}^{2}|^{\frac{1}{2}}$
    & $|{\rm Im}(\delta^{d}_{12})_{LL}(\delta^{d}_{12})_{RR}|^{\frac{1}{2}}$\\
    \noalign{\smallskip} \hline \noalign{\smallskip}
    & 0.3 & 0.0015 & 0.00063 & 0.00020 \\
    & 1.0 & 0.0032 & 0.00035 & 0.00022 \\
    & 4.0 & 0.0075 & 0.00042 & 0.00032 \\
    \noalign{\smallskip} \hline \noalign{\smallskip}
    $\epsilon'/\epsilon$ & $x$ 
    & $|{\rm Im}(\delta^{d}_{12})_{LL}|$
    & \multicolumn{2}{c}{$|{\rm Im}(\delta^{d}_{12})_{LR}|$} \\
    \noalign{\smallskip} \hline \noalign{\smallskip}
    & 0.3 & 0.10 & \multicolumn{2}{c}{0.000011} \\
    & 1.0 & 0.50 & \multicolumn{2}{c}{0.000021} \\
    & 4.1 & 0.27 & \multicolumn{2}{c}{0.000065} \\
    \noalign{\smallskip} 
  \end{tabular}
\end{table}


\begin{thebibliography}{99}
\bibitem{foot} It, however, leaves a possibility that there is 
  some cancellation in $\epsilon$ between the gluino loop discussed here 
  and other possible contributions such as chargino loop to allow a 
  larger supersymmetric contribution to $\epsilon'$.
\end{thebibliography}
\end{document}